\documentstyle[12pt]{article}
\newcommand{\ltwid}{\raise.3ex\hbox{$<$\kern-.75em\lower1ex\hbox{$\sim
$}}}

\begin{document}
\begin{center}
{\large \bf Magnetic Catalysis of Chiral Symmetry Breaking in Gauge Theories.
}
\\[6mm]

V.~P.~Gusynin\\
{\it Bogolyubov Institute for Theoretical Physics,
 Kiev-143, 03143 Ukraine}\\
[8mm]
\end{center}

\begin{abstract}
Non-perturbative effect of the formation of a chiral symmetry breaking
condensate $\langle\bar\psi\psi\rangle$ and of a dynamically generated 
fermion mass in QED in the presence of an external magnetic field is 
considered. The dynamical mass of a fermion (energy
gap in the fermion spectrum) is shown to depend essentially nonanalytically 
on the renormalized coupling constant $\alpha$ in a strong magnetic field.
Possible applications of this effect are discussed.
\end{abstract}
\vskip3mm

The dynamics of fermions in a strong external magnetic field has been 
attracting much attention during last years. Perhaps, the brightest example 
has been the discovery and theoretical explanation of the fractional Hall 
effect leading to the 1998 Nobel Prize award (see Nobel lectures by Laughlin, 
Stormer and Tsui in \cite{nobelprize}) for "discovery that electrons acting 
together in strong magnetic fields can form new types of "particles", with 
charges that are fractions of electron charges ", 
as is said in press release of the Royal Swedish Academy of
Sciencies. Thus, strong magnetic fields can drastically affect the ground 
state of a system leading to new types of excitations.

In this talk, I will describe one more phenomenon in an external magnetic 
field: dynamical breaking of chiral symmetry induced by such a field, hence 
the name magnetic catalysis.(The talk is based on a series of recent papers 
with V.~Miransky and I.~Shovkovy.) This effect has been established as a 
universal phenomenon in $2+1$ and $3+1$ dimensions: a constant magnetic field 
leads to the generation of a fermion dynamical mass at the weakest attractive 
interaction between fermions \cite{GMS0,GMS1,GMS,NP}. The essence of this 
effect is that electrons behave effectively as (1+1)-dimensional ones when 
their energy is much less than the Landau gap $\sqrt{|eB|}$ ($B$ is the 
magnitude of the magnetic field). The lowest Landau level (LLL) 
plays here the role similar to that of the Fermi surface in the BCS theory of 
superconductivity, leading to the dimensional reduction $D\rightarrow D-2$ in 
the dynamics of fermion pairing in a magnetic field and to the formation of 
a chiral condensate at weak coupling. The effect may have interesting 
applications in cosmology \cite{GMS0,Incera} and in condensed matter 
physics \cite{condmat}, as will be discussed below.

The effect of magnetic catalysis was studied in Nambu-Jona-Lasino (NJL)
models in 2+1 \cite{GMS0,Klim} and 3+1 dimensions \cite{GMS1,FGI}, it was
extended to the case of external non-abelian chromomagnetic fields \cite
{nonabelian}, finite temperatures \cite{temperature} and chemical potential
\cite{chempotential,qedchempot}, curved spacetime 
\cite{Odintsov}, confirming the universality of the phenomenon.
In particular, this phenomenon was considered in ($3+1$)-dimensional QED
\cite{GMS,NP,Ng,Hong1,Ferrer,Smilga}. 

We emphasize that we will consider the conventional, weak coupling, phase of
QED since the dynamics of the LLL is long-range (infrared), and the QED 
coupling constant is weak in the infrared region, therefore, the
treatment of the nonperturbative dynamics is reliable there. Note that  chiral 
symmetry breaking is not manifested in the weak coupling phase
of QED in the absense of a magnetic field, even if it is treated 
nonperturbatively \cite{FGM}. We will show that a constant magnetic field $B$
 changes the situation drastically, namely, it leads to dynamical chiral 
symmetry breaking in QED for any arbitrary weak interaction.

The Lagrangian density of  QED in a magnetic field is
\begin{equation}
{\cal L} = -{1\over4}F^{\mu\nu}F_{\mu\nu} + \frac{1}{2} \left[\bar{\psi},
(i\gamma^\mu D_\mu-m_0)\psi\right],
\label{eq:lag}
\end{equation}
where the covariant derivative $D_\mu$ is
\begin{equation}
D_\mu=\partial_\mu-ie(A^{ext}_\mu+A_\mu),\qquad A^{ext}_\mu =
\left(0,-\frac{B}{2} x_2 ,\frac{B}{2} x_1,0\right),
\label{eq:vecA}
\end{equation}
i.e. we use the so-called symmetric gauge for $A_\mu^{ext}$. Besides the Dirac
 index, the fermion field carries an additional flavor index $a=1,2,\dots,N$.
When the bare mass $m_0=0$ the Lagrangian density (\ref{eq:lag}) is  invariant
 under the chiral $SU_L(N)\times SU_R(N)\times U_V(1)$ symmetry
(we will not discuss the dynamics related to the anomalous symmetry 
$U_A(1)$). 

We consider first the problem of free relativistic fermions
in a magnetic field in $3+1$ dimensions and compare it with the same problem
in $2+1$ dimensions. We will see that the roots of the fact that a magnetic
field is a catalyst of chiral symmetry breaking are actually in this dynamics.

The energy spectrum of fermions in a constant magnetic field is:
\begin{equation}
E_n(p_3)=\pm\sqrt{m^2_0+2|eB|n+p_3^2},\quad n=0,1,2\dots
\end{equation}
(the Landau levels). Each Landau level is infinitely degenerate. As the fermion
mass $m_0$ goes to zero, there is no energy gap between the vacuum and the 
lowest Landau level with $n=0$. The density of states of fermions on the 
energy surface with $E_0=0$ is given by 
\begin{equation}
\nu_0=\frac{|eB|N}{4\pi^2},\quad 3+1\quad {\mbox{\rm dimensions,\, and}}
\quad \nu_0=\frac{|eB|N}{2\pi},\quad 2+1\quad{\mbox{\rm dimensions}}.
\end{equation} 
The dynamics of the LLL plays the crucial role in catalyzing spontaneous 
breaking of chiral symmetry. In particular, the density $\nu_0$ plays the 
same role here as the density of states on the Fermi surface $\nu_F$ in the 
theory of superconductivity. The next important point is that the dynamics of 
the LLL is essentially ($1+1$)-dimensional. Indeed, let us consider the fermion
propagator in a magnetic field which was calculated by Schwinger \cite{Sch} 
long ago and has the following form (in the chosen gauge):
\begin{equation}
S(x,y) = \exp \left(\frac{ie}{2}(x-y)^\mu A_\mu^{ext}
(x+y)\right)\tilde{S}(x-y),
\label{barefermprop}
\end{equation}
where the Fourier transform of $\tilde{S}$ is
\begin{eqnarray}
& &\tilde{S}(p) =\int\limits^\infty_0 ds \exp \left[is\left(p^2_0-p^2_3
-p^2_{\perp}\frac{\tan(eBs)}{eBs}-m_{0}^2\right)\right] \nonumber \\
& &\cdot\left[(p^0\gamma^0-p^3\gamma^3+m_{0})(1+\gamma^1\gamma^2\tan(eBs))
-p_{\perp}\gamma_{\perp}(1+\tan^2(eBs))\right].
\label{free_S}
\end{eqnarray}
Here $p_{\perp}=(p_1,p_2)$, $\gamma_{\perp}=(\gamma_1,\gamma_2)$ (to get 
an expression in $2+1$ dimensions we should put $p_3=0$ in (\ref{free_S})).
The  propagator $\tilde{S}(p)$ can be decomposed
over the Landau level poles as follows \cite{Cho} :
\begin{equation}
\tilde{S}(p)=i \exp\left(-\frac{p^2_{\perp}}{|eB|}\right)
\sum_{n=0}^{\infty}(-1)^n\frac{D_n(eB,p)}{p_0^2-p_3^2-m_{0}^2-2|eB|n}
\label{eq:poles}
\end{equation}
with
\begin{eqnarray*}
D_n(eB,p)&=&(\hat{p}_\parallel+m_{0})\left[(1-i\gamma^1\gamma^2
{\rm sign}(eB))L_n\left(2\frac{p^2_{\perp}}{|eB|}\right)\right.\\
&-&\left.(1+i\gamma^1\gamma^2{\rm sign}(eB))
L_{n-1}\left(2\frac{p^2_{\perp}}{|eB|}\right)\right]
+4{\vec p}_\perp{\vec\gamma}_\perp L_{n-1}^1\left(2\frac{p^2_{\perp}}{|eB|}
\right),
\end{eqnarray*}
where $L_n(x)$ are the generalized Laguerre polynomials ($L_n\equiv L_n^0$,
$L_{-1}^{\alpha}(x)=0$).  Eq.(\ref{eq:poles})
implies that at $p^2_{\parallel}, m_{0}^2\ll \sqrt{|eB|}$, the LLL with $n=0$
dominates and we can write
\begin{eqnarray}
\tilde{S}(p)\simeq 2i\exp(-\frac{p^2_{\perp}}{|eB|})\frac{\hat{p}_{\parallel}
+m_{0}}{p^2_{\parallel}-m^2_{0}}O^{(-)},
\label{eq:tildeS}
\end{eqnarray}
where $\hat{p}_{\parallel}=p^0\gamma^0-p^3\gamma^3$ and
$\hat{p}_{\parallel}^2=(p^0)^2-(p^3)^2$. The matrix $O^{(-)}\equiv(1-i\gamma^1
\gamma^2{\rm sign}(eB))/2$ is the
projection operator on the fermion states with the spin polarized
along the magnetic field.
This point and Eq.~(\ref{eq:tildeS})
clearly demonstrate the (1+1)-dimensional character of the dynamics
of fermions in the LLL. This property is preserved also in the
case when the fermion mass is generated dynamically. Since at $m_0^2,p^2_\parallel,p^2_\perp\ll|eB|$ the LLL pole dominates in the fermion propagator,
 one concludes that the
dimensional reduction ($D\rightarrow D-2$) takes place for the infrared 
dynamics in a strong ($|eB|>>m_0^2$) magnetic field. Such a dimensional 
reduction reflects the fact that the motion of charged particles is restricted
 in directions perpendicular to the magnetic field.

Let us first calculate the chiral condensate in $2+1$ dimensions for free 
four-component fermions:
\begin{eqnarray}
&&\langle0|\bar\psi\psi|0\rangle=-\lim_{x\to y}{\rm tr}S(x,y)=-\frac{i}
{(2\pi)^3}{\rm tr}\int d^3p\tilde{S}_E(p)\nonumber\\
&&= -\frac{4m_0N}{(2\pi)^3}\int d^3p
\int_{1/\Lambda^2}^\infty ds\exp\left[-s\left(m_0^2+p_3^2+{\vec p}^2_\perp\frac
{\tanh(eBs)}{eBs}\right)\right]\nonumber\\
&&=-m_0\frac{|eB|N}{2\pi^{3/2}}\int_{1/\Lambda^2}^\infty\frac{ds}{s^{1/2}}
e^{-sm_0^2}\coth\left(|eBs|\right)\rightarrow-\frac{|eB|N}{2\pi},\quad m_0\to0,
\label{2+1cond}
\end{eqnarray}
where $\Lambda$ is an ultraviolet cutoff in Euclidean space and, for 
concretness, we consider $m_0\ge0$. Thus, as $m_0\to0$, 
the condensate $\langle0|\bar\psi\psi|0\rangle$ remains non-zero due 
to the LLL. Note that the expression (\ref{2+1cond}) is nothing else as
the Banks-Casher formula relating the fermion condensate to the level
density of the Dirac operator at zero eigenvalue \cite{Banks}. The appearance 
of the condensate in the chiral (flavor) limit, $m_0\to0$, signals the 
spontaneous breakdown of the chiral (flavor) symmetry even for free fermions in
a magnetic field at $D=2+1$ \cite{GMS0}. 

Repeating the same calculation of the chiral condensate in $3+1$ dimensions
 we would get
\begin{equation}
\langle0|\bar\psi\psi|0\rangle\simeq-m_0\frac{|eB|N}{4\pi^2}\left(\ln\frac
{\Lambda^2}{m_0^2}+O(1)\right),\qquad m_0\to0,
\label{3+1cond}
\end{equation}
i.e. the condensate is zero and there is  no chiral symmetry breaking.
Note, however, the appearance of logarithmic singularity in (\ref{3+1cond}) 
due to the LLL dynamics. As we will see below, switching on even a weak 
attraction between fermions leads to the formation of chiral condensate in
($3+1$)-dimensional case. 

The above consideration suggests that there is a universal mechanism for
enhancing the generation of fermion masses by a strong magnetic field:
the fermion pairing takes place essentially for fermions at the LLL and
this pairing dynamics is $(1+1)$-dimensional in the infrared region.This 
is the main reason why in a magnetic field spontaneous chiral symmetry
breaking takes place even at the weakest attractive interaction between 
fermions in $3+1$ dimensions \cite{GMS1,GMS,NP}.

Now we shall consider QED in $3+1$ dimensions whose Lagrangian is given by Eq.
(\ref{eq:lag}). To study chiral symmetry breaking one has to solve the
Schwinger-Dyson (SD) equation for the dynamical fermion mass. The SD
equation for the fermion propagator $G(x,y)$ in an external field has the form
\begin{eqnarray}
&&G^{-1}(x,y)=S^{-1}(x,y)+\Sigma(x,y),\label{G{-1}}\\
&&\Sigma(x,y)=4\pi\alpha\gamma^\mu\int G(x,z)
\Gamma^\nu (z,y,z^\prime){\cal D}_{\nu\mu}
(z^\prime,x)d^4zd^4z^\prime.
\label{SD-fer}
\end{eqnarray}
Here $S(x,y)$ is the bare fermion propagator (\ref{barefermprop}) in 
the external field $A_{\mu}^{ext}$, $\Sigma(x,y)$ is the fermion mass
operator, and ${\cal D}_{\mu\nu}(x,y)$, $\Gamma^\nu (x,y,z)$ are
the full photon propagator and the full amputated vertex.
The full photon propagator satisfies the equations
\begin{eqnarray}
{\cal D}^{-1}_{\mu\nu}(x,y)&=&D^{-1}_{\mu\nu}(x-y)
+\Pi_{\mu\nu}(x,y), \label{SD-pho}\\
\Pi_{\mu\nu}(x,y)&=&-4\pi\alpha \mbox{tr} \gamma_{\mu}
\int d^4 u d^4 z G(x,u)\Gamma_{\nu} (u,z,y) G(z,x),
\label{Pi_munu}
\end{eqnarray}
where $D_{\mu\nu}(x-y)$ is the free photon propagator and
$\Pi_{\mu\nu}(x,y)$ is the polarization operator.

It is not difficult to show directly from the SD equations (\ref{G{-1}}),
(\ref{SD-fer}), (\ref{SD-pho}) and (\ref{Pi_munu}) that substitutions
%\begin{mathletters}
\begin{eqnarray}
&&G(x,y)=\exp\left(ie x^{\mu} A^{ext}_{\mu}(y)\right)
\tilde{G}(x-y),
\label{14a}\\
&&\Gamma(x,y,z)=\exp\left(ie x^{\mu} A^{ext}_{\mu}(y)\right)
\tilde{\Gamma}(x-z,y-z),
\label{14b}\\
&&{\cal D}_{\mu\nu}(x,y)=\tilde{{\cal D}}_{\mu\nu}(x-y),
\label{14c}\\
&&\Pi_{\mu\nu}(x,y)=\tilde{\Pi}_{\mu\nu}(x-y)
\label{14d}
\end{eqnarray}
%\label{14}
%\end{mathletters}
\noindent
lead to equations for translation invariant parts of Green's functions.
In other words, in a constant magnetic field, the Schwinger phase
is universal for Green's functions containing one fermion field,
one antifermion field, and any number of photon fields, and the
full photon propagator is translation invariant.

We solve the SD equation for the fermion propagator in the so-called  
ladder approximation when the full vertex and full photon propagator
are replaced by their bare ones. We have
\begin{equation}
\tilde{G}(x)=\tilde{S}(x)
-4\pi \alpha \hspace{-1mm}\int \hspace{-1mm}d^4 x_1 d^4 y_1 e^{ix A(x_1)
+ix_1 A(y_1)}\tilde{S}(x-x_1)\gamma^{\mu}\tilde{G}(x_1-y_1)\gamma^{\nu}
\tilde{G}(y_1) {\cal D}_{\mu\nu}(x_1-y_1),
\end{equation}
where the shorthand $xA^{ext}(y)$ stands for $x^\mu A_\mu^{ext}(y)$.

First, let us show that the solution to the above equation,
$\tilde{G}(x)$, allows the factorization of the dependence on 
the parallel and perpendicular coordinates,
\begin{equation}
\tilde{G}(x)=\frac{i}{2\pi l^2}
\exp\left(-\frac{x_{\perp}^2}{4l^2}\right)
g\left(x_{\parallel}\right)O^{(-)}.
\label{B4}
\end{equation}
Notice that this form for $\tilde{G}(x)$ is suggested by a
similar expression for the bare propagator,
\begin{equation}
\tilde{S}(x)=\frac{i}{2\pi l^2}
\exp\left(-\frac{x_{\perp}^2}{4l^2}\right)
s\left(x_{\parallel}\right)O^{(-)},
\label{B5}
\end{equation}
with
\begin{equation}
s\left(x_{\parallel}\right)
=\int\frac{d^2 k_{\parallel}}{(2\pi)^2}
e^{-ik_{\parallel}x_{\parallel}}
\frac{\hat{k}_{\parallel}+m}{k_{\parallel}^2-m^2},
\label{B6}
\end{equation}
taken in the LLL approximation (here $l=|eB|^{-1/2}$ is the magnetic length).
Performing the integrations we arrive at
\begin{eqnarray}
g\left(x_{\parallel}\right)=s\left(x_{\parallel}\right)
&+&4\pi \alpha \int \frac{d^4 q}{(2\pi)^4}
d^2 x_1^{\parallel} d^2 y_1^{\parallel}
\exp\left(-\frac{(q_{\perp}l)^2}{2}
-iq_{\parallel}(x_1^{\parallel}-y_1^{\parallel})\right)
s(x^{\parallel}-x_1^{\parallel})\nonumber\\
&\times&\gamma^{\mu}_{\parallel}
g(x_1^{\parallel}-y_1^{\parallel})\gamma^{\nu}_{\parallel}
g(y_1^{\parallel})
{\cal D}_{\mu\nu}\left(q_{\parallel},q_{\perp}\right).
\end{eqnarray}
Regarding this equation, it is necessary to emphasize that the
``perpendicular" components of the $\gamma$-matrices are absent
in it. Indeed, because of the identity  $O^{(-)}
\gamma_{\perp}^{\mu}O^{(-)}=0$, all those components are killed
by the projection operators coming from the fermion propagators.
By switching to the momentum space, we obtain
\begin{eqnarray}
g^{-1}\left(p_{\parallel}\right)
=s^{-1}\left(p_{\parallel}\right)
-4\pi \alpha \int \frac{d^4 q}{(2\pi)^4}
\exp\left(-\frac{(q_{\perp}l)^2}{2}\right)
\gamma^{\mu}_{\parallel}
g(p^{\parallel}-q^{\parallel})\gamma^{\nu}_{\parallel}
{\cal D}_{\mu\nu}\left(q_{\parallel},q_{\perp}\right).
\end{eqnarray}
The general solution to this equation is given by the ansatz,
\begin{equation}
g\left(p_{\parallel}\right)=\frac{A\hat{p}_{\parallel}+B}
{A^2 p_{\parallel}^2-B^2},
\label{B12}
\end{equation}
where $A$ and $B$ are functions of $p_{\parallel}^2$.
Making use of this as well as of the explicit form of the photon
propagator in the Feynman gauge, we get that the function $A=1$ 
while for the mass function we get the following integral equation
\begin{equation}
B(p^2)=m_0+\frac{\alpha}{2\pi^2}
\int\frac{d^2q B(q^2)} {q^2+B^2(q^2)}
\int\limits_{0}^{\infty}\frac{dx \exp(-xl^2/2)}{x+( q- p)^2}
\label{inteq:B}
\end{equation}
(henceforth we will omit the symbol $\parallel$ from $p$ and $q$).
Thus the SD equation has been reduced to a two--dimensional integral
equation. Of course, this fact reflects the two--dimensional character of
the dynamics of electrons from LLL. 

Analytical and numerical analysis of this equation were performed in
\cite{NP} for the case $m_0=0$ and in \cite{Smilga} for nonzero bare mass.
The numerical analysis showed that the so called linearized approximation,
 with $B(q^2)$ replaced by the total mass
$m_{\rm tot}\equiv B(0)$ in the denominator of Eq.~(\ref{inteq:B}),
is an excellent approximation. Then we get
\begin{equation}
B(p^2)=m_0+\frac{\alpha}{2\pi^2}
\int\frac{d^2q B(q^2)} {q^2+m_{tot}^2}
\int\limits_{0}^{\infty}\frac{dx \exp(-xl^2/2)}
{x+( q- p)^2}.
\label{integlinear}
\end{equation}

As was shown in \cite{NP} (see Appendix C), in the case of weak coupling 
$\alpha$ and for $m_0=0$, the function $B(p)$ remains almost
constant in the range of momenta $0<p^2\ltwid 1/l^2$ and decays like $1/p^2$
outside that region. To get an estimate for $m_{dyn}\equiv B(0)$ at 
$\alpha<<1$, we set
the external momentum to be zero and notice that the main contribution in the
integral on the right hand side of Eq.(\ref{integlinear})
is formed in the infrared region with $q^2\ltwid 1/l^2$. The latter validates
in its turn the substitution $B(q)\rightarrow B(0)$ in the integrand of (\ref
{inteq:B}), and we finally come to the following gap equation:
\begin{equation}
B(0)\simeq \frac{\alpha}{2\pi^2}B(0)\int\frac{d^2q}{q^2+m^2_{dyn}}
\int\limits^{\infty}_{0} \frac{dx \exp(-l^2x/2)}{q^2+x},
\label{gapeq}
\end{equation}
which gives the expression for the dynamical fermion mass
(energy gap in the fermion spectrum):
\begin{equation}
m_{dyn}\simeq C\sqrt{eB}\exp{\left[-\sqrt{\frac{\pi}{\alpha}}
\right]},
\label{mdyn}
\end{equation}
where the constant $C$ is of order one and $\alpha=e^2 /4\pi$ is the
renormalized coupling constant related to the scale $\sqrt{eB}$.
The exponential factor in $m_{dyn}$ displays the nonperturbative nature of 
this result. It can be shown also that the expression (\ref{mdyn}) for the 
dynamical mass is gauge invariant \cite{GMS}.

A more accurate analysis \cite{NP}, which takes into account the momentum
dependence of the mass function, leads to the result
\begin{equation}
m_{dyn}\simeq C\sqrt{|eB|}\exp\left[-{\pi\over2}\sqrt{\frac{\pi}{2\alpha}}
\right].
\label{mdynmomdep}
\end{equation}
The ratio of the powers of this exponent and that in Eq.(\ref
{mdyn}) is $\pi/2\sqrt2\simeq1.1$, thus the approximation used above is
rather reliable. 

We note that $m_{dyn}$ has rather unusual $1/\sqrt\alpha$
behavior of the exponents in (\ref{mdyn}) and (\ref{mdynmomdep}). Similar 
dependence was found recently in QCD for a quark gap arising at high 
densities (color superconductivity) \cite{Son}.
The reason for such a behavior in both cases is the same: the presence of 
long-range interactions.

To study chiral symmetry breaking in an external field at nonzero temperature
we use the imaginary time formalism. Now the analogue of the equation 
(\ref{integlinear}) (with $m_0=0$ and the replacement $m_{dyn}\rightarrow 
m^2(T)$ in the denominator) reads
\begin{equation}
B(\omega_{n'},p)=\frac{\alpha}{\pi}T\sum_{n=-\infty}^\infty\int\limits_
{-\infty}^\infty\frac{dkB(\omega_n,k)}{\omega^2_n+k^2+m^2(T)}\int\limits_0^
\infty\frac{dx\exp(-l^2x/2)}{(\omega_n-\omega_{n'})^2+(k-p)^2+x},
\label{tempgap}
\end{equation}
where $\omega_n=\pi T(2n+1)$ are Matsubara frequencies.

If we now take $n'=0, p=0$ in the left hand side of Eq.(\ref{tempgap}) and
put $B(\omega_n,k)\approx B(\omega_0,0)=const$ in the integrand, we come to
the equation
\begin{equation}
1=\frac{\alpha}{\pi}T\sum_{n=-\infty}^\infty\int\limits_{-\infty}^\infty
\frac{dk}{\omega^2_n+k^2+m^2(T)}\int\limits_0^\infty\frac{dx\exp(-l^2x/2)}
{(\omega_n-\omega_0)^2+k^2+x}.
\label{tgap}
\end{equation}
The equation for the critical temperature is obtained
 putting $m(T_c)=0$ and this determines the critical temperature \cite
{temperature}
\begin{eqnarray}
T_{c}\approx \sqrt{|eB|}\exp\left[-\sqrt{\frac{\pi}{\alpha}}\right]
\approx m_{dyn}(T=0),
\label{t_c-m_dyn}
\end{eqnarray}
where $m_{dyn}$ is given by (\ref{mdyn}). The relationship
$T_c\approx m_{dyn}$ between the critical temperature and the zero
temperature fermion mass was obtained also in NJL model in (2+1)-
and (3+1)-dimensions (\cite{GMS0} and second paper in Ref.\cite{nonabelian}). 
The constant $C$, in the
relation $T_c=Cm_{dyn}$, is of order one and can be calculated
numerically.  We note that the photon thermal mass, which is of the order
of $\sqrt\alpha T$ \cite{Weldon}, cannot change our result for the
critical temperature.

Taking into account the non-zero bare electron mass we come to the equation
for the total mass $m$:
\begin{equation}
m\cos\left(\sqrt{\frac{\alpha}{2\pi}}\log\frac{|eB|}{m^2}\right)=m_0.
\label{totalmass}
\end{equation}
It can be shown \cite{Smilga} that the itterative solution of last equation
reproduces all leading double logarithmic terms in perturbation theory:
\begin{equation}
m=m_0\left[1+\frac{\alpha}{4\pi}\log^2\frac{|eB|}{m_0^2}+\frac{5}{24}
\left(\frac{\alpha}{2\pi}\log^2\frac{|eB|}{m_0^2}\right)^2+\frac{61}{720}
\left(\frac{\alpha}{2\pi}\log^2\frac{|eB|}{m_0^2}\right)^3+\cdots. 
\right]
\label{etaexpansion}
\end{equation}
From Eq.(\ref{totalmass}) we can estimate the dynamical mass due to a magnetic
 field. For fields of the order of $\sim10^{14} G$ which are realized on 
surfaces of neutron stars we get ${(m-m_0)}/m \sim 10\%$. In the 
real QED the expansion parameter $\eta\equiv\frac{\alpha}{2\pi}\log^2
(|eB|/m_0^2)\sim1$ in 
(\ref{etaexpansion}) explores the transition between the perturbative regime
$\eta\ll1$ and the nonperturbative massless QED regime $\eta\gg1$. The
value of the parameter $\eta\simeq1$ is reached at fields of the order
$\sim10^{26} G$. We recall that strong magnetic fields ($B\sim10^{24} G$)
might have been generated during the electroweak phase transition \cite
{EWfields}. It has been speculated in Refs.\cite{GMS,NP} that the character 
of electroweak phase transition could be affected by a generation of a 
dynamical electron mass under such strong fields. The nonperturbative regime
becomes prevailing over the perturbative one for values of $\eta$ of the
order of $2.35$ \cite{Smilga} what corresponds to magnetic fields 
$\sim10^{32} G$. Ambj{\o}rn and Olesen \cite{Olesen} have claimed that even 
larger fields, $\sim10^{33} G$, would be necessary at early stages of the 
Universe to explain the observed large-scale galactic magnetic fields. 

Since the induced fermion dynamical mass contains an exponential factor
(see (\ref{mdyn}), (\ref{mdynmomdep}) ) it is quite small at all reasonable
values of the coupling $\alpha$, therefore, there are tiny chances to find
implications of the magnetic catalysis phenomenon in real experiments. However,
it was shown recently \cite{FerrerIncera} that the Yukawa coupling and
scalar-scalar interaction can considerably enhance the fermion dynamical mass 
(according to \cite{FerrerIncera} the dynamical mass is estimated to be 
$m_{dyn}\simeq 0.6\sqrt{|eB|}$). The most immediate physical implication would 
be then in the electroweak theory.

Another interesting application of the magnetic catalysis phenomenon is found
in ($2+1$)-dimensional condensed matter systems \cite{condmat}, given the 
suggestions that high-temperature superconductors can be described effectively
by ($2+1$) relativistic field theories like NJL or QED (the relativistic 
(Dirac) nature of the fermion fields is related to the fact that they describe
 the quasi-particle excitations about the nodes of a $d$-wave superconducting 
gap). According to recent experiments \cite{Krishana}, at temperatures 
significantly lower than $T_c$ of superconductivity, the thermal 
conductivity, as a function of a magnetic field perpendicularly applied to 
the cuprate planes, displays a sharp break in its slope at a transition field
$B_\kappa$, followed then by  a plateaux region in which it ceases to change 
with increasing field. 
The critical temperature for appearance of the kink-like
 behavior scales with the magnetic field as $T_\kappa\sim\sqrt{e|B|}$. This 
phenomenon may indicate the opening of a second gap, at the nodes of the 
$d$-wave superconducting gap, that depends on the strength of the applied
magnetic field \cite{condmat,Krishana,Laughlin}. Indeed, as we saw, in  
($2+1$)-dimensional systems the chiral condensate appears even in absence of 
interaction between fermions. The dynamically generated fermion mass 
scales with a magnetic field like $m_{dyn}\sim \sqrt{e|B|}$ in $2+1$ NJL 
model \cite{GMS0}, and $m_{dyn}\sim\alpha\log({\sqrt{|eB|}/\alpha})$ in 
QED3 \cite{Shpagin}. The critical temperature for vanishing of the
dynamical mass is determined by the dynamical mass at zero temperature (see 
Eq.(\ref{t_c-m_dyn})) and scales with a magnetic field in a way quite similar 
to the scaling law found in experiments.  

In conclusion, we discuss very breifly the role of higher order radiative
corrections in the magnetic catalysis problem. As was shown in Ref.\cite{NP},
because of the (1+1)-dimensional form of the fermion propagator of the LLL
fermions, there are relevant higher order contributions. In particular, 
considering this problem in the improved rainbow approximation (with the 
bare vertex in the Schwinger-Dyson equations for both the fermion propagator
 and the polarization operator ), it was shown that the fermion mass is given 
by Eq. (\ref{mdynmomdep}) but with $\alpha\rightarrow\alpha/2$. Recently we 
have shown \cite{PRLNucl} that there exists a special (non-local) gauge in 
which the SD equations written in the improved rainbow approximation 
are reliable: in other words, in that gauge there exists a consistent 
truncation of the Schwinger-Dyson equations for this non-perturbative problem.
  The expression for $m_{dyn}$ takes the following form,
\begin{equation}
m_{dyn} =\tilde C \sqrt{|eB|} F(\alpha)\exp\left[-\frac{\pi}
{\alpha\ln\left(C_1/N\alpha\right)}\right], \label{m}
\end{equation}
where $N$ is the number of fermion flavors, $F(\alpha) \simeq
(N\alpha)^{1/3}$, $C_1\simeq 1.82$ and the constant $\tilde C$ is of order one.

Thus, the magnetic catalysis of chiral symmetry breaking in QED 
yields a (first, to the best of our knowledge) example in which
there exists a consistent truncation of the Schwinger-Dyson
equations in the problem of dynamical symmetry breaking in a
(3+1)-dimensional gauge theory without fundamental scalar fields.
\vskip3mm
%\vspace*{-2pt}
%\section*{Acknowledgments}
I am grateful to the members of the Institute for Theoretical Physics of
the University of Heidelberg, especially Prof. M.G. Schmidt, for their
hospitality during my stay there.
This research has been supported in part by Deutscher Academischer 
Austauschdienst (DAAD) grant and by the National Science Foundation (USA) under
grant No. PHY-9722059.

%\vspace*{-9pt}

\end{document}